\documentstyle[floats,prb,aps,epsf,twocolumn]{revtex}
\begin{document}
\draft
\title{Anomalous proximity effect in $d$-wave superconductors}
\author{A.\ A.\ Golubov$^{1,2}$, M.\ Yu.\ Kupriyanov $^{1,3}$}
\address{$1)$ Department of Applied Physics, University of Twente,
P.O.Box 217 \\ 7500 AE Enschede, The Netherlands\\
$2)$ Institute of Solid Physics, 142432 Chernogolovka, Russia\\
$3)$ Nuclear Physics Institute, Moscow State University, 119899 GSP \\
Moscow, Russia}
\maketitle

\begin{abstract}
The anomalous proximity effect between a $d$-wave superconductor and a surface 
layer with small electronic mean free path is studied theoretically in the 
framework of the Eilenberger equations. The angular and spatial structure 
of the pair potential and the quasiclassical propagators in the interface 
region is calculated selfconsistently. The variation of the spatially-resolved 
quasiparticle density of states from the bulk to the surface is studied. 
It is shown that the isotropic gapless superconducting state is induced 
in the disordered layer.
\end{abstract}



\section{Introduction}

There is continuing experimental evidence that the behavior of high
temperature superconductors (HTS) can be understood in terms of the $d$-wave
pairing scenario, rather than in the conventional $s$-wave picture. On the
other hand it is well known that the $d$-wave order parameter is strongly
reduced by electron scattering at impurities and therefore can be formed
only in clean materials. However, the condition of clean limit is not
fulfilled in the vicinity of the grain boundaries or other HTS interfaces
even if the material is clean in the bulk. There are at least two reasons for
that. The first one is that quasiparticle reflection from realistic
interfaces is diffusive, thus providing the isotropisation in the momentum
space and the suppression of the $d$-wave component of the order parameter.
The second one is the contamination of the material near interfaces as a
result of fabrication process or electromigration in large scale
application devices. As a result the formation of a thin disordered layer
near HTS surfaces and interfaces is highly probable. An important question
is whether or not superconducting correlations vanish in such a layer
in the limit of small mean free path and what is the orbital structure of the
superconducting state in the interface region.

Surface peculiarities in $d$-wave superconductors were extensively discussed in the
framework of the theoretical models based on specular quasiparticle reflection from
clean interfaces \cite{Hu,Kashiwaya,Tan1,Tan2,Barash1,Rainer1,Rainer2}.
Zero- and finite-bias anomalies predicted in these papers were recently
observed experimentally in Refs. \cite{Alf,Greene}. In this paper we focus
on the problem of the anomalous proximity effect between a $d$-wave superconductor 
and a thin disordered layer in the limit of strong disorder. It is shown that an 
isotropic order parameter is nucleated in such a layer even in the absence of the 
subdominant pairing interaction in the $s$-wave channel. The spatially-resolved 
quasiparticle density of states is calculated. It is shown that zero- and 
finite-energy peaks are present in the surface density of states in the $d$-wave 
region. Zero-energy peaks are fully smeared out in the disordered layer, 
which is in a peculiar gapless superconducting state.

\section{Proximity effect at the interface}

Two approaches to the study of surface roughness effects in unconventional 
superconductors were used previously. In the first one it is assumed that 
the interface consists of facets with random orientations compared to the
crystallographic axes of the material \cite{Rainer1}. According to the second approach,
both sides of an ideal interface are coated by a so-called Ovchinnikov's thin disordered
layer \cite{Barash1,Rainer3,Ovch}. In the latter case the degree of disorder
(or interface roughness) is measured by the ratio of the layer thickness $d$
to the quasiparticle mean free path in the layer $\ell.$ 
Up to now both approaches were used to
study the smearing of Andreev surface bound states by weak
disorder. Here we will concentrate on the regime of strong disorder.

We consider the surface or a weakly transparent barrier in a $d$-wave superconductor 
oriented normal to the crystallographic $ab$ plane. 
We assume that the crossover from the clean to the dirty limit
takes place in a thin layer near the surface with mean free path $\ell $
and thickness $d<\sqrt{\xi _0\ell }$, where $\xi _0$ is the coherence
length of the bulk material.

To study the proximity effect at the interface we use the quasiclassical
Eilenberger equations \cite{Eilenb} with impurity scattering taken in the Born
limit. For our purpose it is convenient to rewrite these equations in
terms of functions $\Phi _{+}=(f(r,\theta )+f(r,\theta +\pi ))/2$ and $\Phi
_{-}=f(r,\theta )-f(r,\theta +\pi )$

\begin{equation}
4\omega \Phi _{+}+v\cos \theta \frac{d\Phi _{-}}{dx}=4\Delta g+\frac 2\tau
(g\left\langle \Phi _{+}\right\rangle -\Phi _{+}\left\langle g\right\rangle )
\label{Eq1}
\end{equation}

\begin{equation}
2v\cos \theta \frac{d\Phi _{+}}{dx}=-(2\omega +\frac 1\tau \left\langle
g\right\rangle )\Phi _{-}  \label{Eq.2}
\end{equation}

\begin{equation}
2v\cos \theta \frac{dg}{dx}=(2\Delta +\frac 1\tau \left\langle \Phi
_{+}\right\rangle )\Phi _{-}  \label{Eq.3}
\end{equation}

\begin{equation}
\Delta \ln \frac T{T_c}+2\pi T\sum_{\omega >0}\left( \frac \Delta \omega
-\left\langle \lambda (\theta ,\theta ^{\prime })\Phi _{+}\right\rangle
\right) =0.  \label{Eq.4}
\end{equation}
Here $g$ and $f$ are respectively the normal and the anomalous quasiclassical 
propagators, $\omega =\pi T(2n+1)$ are the Matsubara frequencies, $v$ is the Fermi
velocity, $x$ is the coordinate in the direction of the surface normal, $%
\theta $ is the angle between the surface normal and quasiparticle
trajectory, $\tau =\ell /v$ and $\left\langle ...\right\rangle =(1/2\pi
)\int_0^{2\pi }(...)d\theta $. We assume that the Fermi surface has a
cylindrical shape. 

For pure $d$-wave interaction the coupling constant
may be written in the form \cite{carbotte}

\[
\lambda (\theta ,\theta ^{\prime })\equiv
\lambda _d(\theta ,\theta ^{\prime })=
2\lambda \cos (2(\theta -\alpha ))\cos (2(\theta^{\prime }-\alpha)),  
\]
where $\alpha $ is the misorientation angle
between the crystallographic $a$ axis and the surface normal.
Then according to the self-consistency equation (\ref{Eq.4}) 
angular and spatial dependencies of the pair potential are 
factorized $\Delta =\sqrt{2}\Delta(x)\cos (2(\theta -\alpha ))$, i.e.
$\Delta$ has pure $d$-wave angular structure everywhere in the interface region.

Far from the interface the bulk anomalous propagator also has the 
$d$-wave symmetry
 
\begin{equation}
\Phi _{+}=\frac{\sqrt{2}\Delta _\infty \cos (2(\theta -\alpha ))}{\sqrt{%
\omega ^2+2\Delta _\infty ^2\cos ^2(2(\theta -\alpha ))}}.  \label{Eq.4a}
\end{equation}
At the same time, as will be shown below, the angular structure of the propagator
$\Phi _{+}(x,\theta )$ is essentially modified near the interface and an $s$-wave
component of $\Phi _{+}(x,\theta )$ is induced.

To proceed further we have to supplement equations (\ref{Eq1})-(\ref{Eq.4})
with the appropriate boundary conditions for the function $\Phi _{+}(x)$\ and its 
derivative $d\Phi _{+}(x)/dx$ at the interface between the clean and the
disordered regions of a $d$-wave superconductor (at $x=0$). These conditions
can be derived by integration of the Eilenberger equations (\ref{Eq1})-(\ref
{Eq.2}) in a small region near the interface. In accordance with Ref.\cite
{Zai}, the first boundary condition is the continuity of $\Phi _{-}$\ at the
interface and can be written in the form 
\begin{equation}
\frac{\ell \cos \theta }{\left\langle g(-0)\right\rangle }\frac{d\Phi
_{+}(-0)}{dx}=\frac{v\cos \theta }{2\omega }\frac{d\Phi _{+}(+0)}{dx}.
\label{Eq.5a}
\end{equation}
This condition manifests the current conservation across the interface.

The second boundary condition depends on the backscattering properties of the
interface. To account for such a backscattering we introduce a strongly
disordered thin layer located near the interface at $-\delta \leq x\leq 0$,
which is characterized by the mean free path $\ell _\delta $, where $\ell
_\delta ,\delta \ll d,\ell $. Assuming that all the boundaries are
transparent, integrating Eq.(\ref{Eq.2}) in the interval $-\delta
\leq x\leq 0$\ \ and taking the limit $\delta \rightarrow 0$\ we arrive at 

\begin{equation}
D\ell \frac{d\Phi _{+}(-0)}{dx}=\Phi _{+}(+0)-\Phi _{+}(-0),  \label{Eq.5b}
\end{equation}
where $D=2\delta /\ell _\delta $. 

For $D=0$\ Eq.(\ref{Eq.5b}) is a
direct consequence of the continuity of the Eilenberger functions along
quasiclassical trajectories which is valid for transparent $SN$-boundaries 
\cite{Zai}. With increase of $D$\ the probability of quasiparticle
penetration into the $N$-layer decreases as $D^{-1}$, i.e. $\Phi _{+}(-0)$\ $%
\approx D^{-1}\Phi _{+}(+0)$. It means that most quasiparticles are
diffusively reflected back to the $d$-wave region at a length scale
smaller than $\ell $.

In the following we will consider the case of strong disorder $\ell \ll d$.
Then in accordance with Refs.\cite{Zai,KL} it follows from the symmetry of 
the problem that the boundary condition at the totally reflecting free 
interface $(x=-d)$ is 
\begin{equation}
\frac d{dx}\Phi _{+}(-d)=0.  \label{Eq.5}
\end{equation}

Since $\ell \ll d$ and $d<\sqrt{\xi _0\ell }$, the dirty limit condition 
$\ell \ll \xi _0$ is fulfilled in the disordered layer. It is straightforward 
to show from Eqs.(\ref{Eq1}), (\ref{Eq.4}) that in this case the pair potential 
$\Delta$ in the disordered layer vanishes due to impurity pair-breaking. 
Then the angle-averaged functions $\left\langle\Phi _{+}\right\rangle $ and 
$\left\langle g\right\rangle $ at $-d\leq x\ll -\ell $ obey
the dirty limit Usadel equations \cite{Usadel} in the form which 
formally coincides with the one valued for a normal metal with 
$T_{cn}=0$. Since the scale of variation of $\left\langle\Phi _{+}\right\rangle $
and $\left\langle g\right\rangle $ in this regime is of the
order of the dirty limit coherence length $\sqrt{\xi _0\ell },$ 
the functions $\left\langle \Phi _{+}\right\rangle $ and $\left\langle g\right\rangle $ 
in a thin disordered layer $d<\sqrt{\xi _0\ell }$ are spatially-independent.

As a result the Eilenberger equations in the region $-d\leq x\leq 0$ are
essentially simplified and have the solution 
\begin{equation}
\Phi _{+}=\left\langle \Phi _{+}\right\rangle +A\frac{\cosh (k(x+d))}{\cosh
(kd)},\quad \Phi _{-}=-2A\frac{\sinh (k(x+d))}{\cosh (kd)},\qquad
\label{Eq.6}
\end{equation}

\begin{equation}
g=\left\langle g\right\rangle -\frac{\left\langle \Phi _{+}\right\rangle }{%
\left\langle g\right\rangle }A\frac{\cosh (k(x+d))}{\cosh (kd)},\qquad
\left\langle \Phi _{+}\right\rangle ^2+\left\langle g\right\rangle ^2=1,
\label{Eq.7}
\end{equation}
where $k=1/\ell \left| \cos \theta \right| .$

Making use of the boundary conditions Eqs.(\ref{Eq.5a}), (\ref{Eq.5b}) at $x=0$ and
of Eqs.(\ref{Eq.6}), (\ref{Eq.7}), one can further reduce the problem to the
solution of the Eilenberger equations (\ref{Eq1})-(\ref{Eq.4}) in the clean $%
d$-wave superconductor $(x\geq 0)$%
\begin{equation}
\kappa ^2\frac{d^2\Phi _{+}}{dx^2}-\Phi _{+}=-\frac \Delta \omega g,\quad
\kappa =\frac{v\left| \cos \theta \right| }{2\omega },  \label{Eq.9}
\end{equation}
\begin{equation}
\frac{dg}{dx}=-\frac \Delta \omega \frac{d\Phi _{+}}{dx}  \label{Eq.10}
\end{equation}
with the condition (\ref{Eq.4a}) in the bulk ($x\gg \xi _0$) and the following 
boundary condition at $x=0$ 
\begin{equation}
\left\{ \kappa \left\langle g(0)\right\rangle +D\frac v\omega \right\} \frac 
d{dx}\Phi _{+}(0)=\Phi _{+}(0)-\left\langle \Phi _{+}(0)\right\rangle .
\label{Eq.11}
\end{equation}
Since
\[
\left\langle g(0)\right\rangle =\sqrt{1-\left\langle \Phi
_{+}(0)\right\rangle ^2},
\]
the boundary condition (\ref{Eq.11}) has a closed form. 

In the following we will limit ourselves to the situation when the
disordered layer produces the most strong effect, namely when $D=0$. 
The isotropic Usadel function $\left\langle \Phi _{+}(0)\right\rangle $
has to be determined selfconsistently by an iteration procedure.

\section{Results and discussion}

In the limit $\kappa \ll \xi _0$ the pair potential $\Delta (x)$ is a smooth
function of $x$ at distances of the order of $\kappa $. Then the boundary
value problem Eqs.(11)-(13) is essentially simplified and has the asymptotic
solution 
\begin{equation}
\Phi _{+}=\Psi (x)+\eta \sqrt{\frac{G(x)}{G(0)}}\exp \left\{ -\int_0^x\frac{%
dy}{\kappa G(y)}\right\} ,\qquad  \label{Eq.12}
\end{equation}
where

\[
\eta =G(0)\frac{\left\langle \Phi _{+}(0)\right\rangle -\Psi
(0)+\left\langle g(0)\right\rangle \kappa \Psi ^{\prime }(0)}{%
G(0)+\left\langle g(0)\right\rangle [1-\kappa (\Psi (0)/\Delta )^{\prime }]}%
, 
\]

\[
\qquad G(x)=\frac \omega {\sqrt{\omega ^2+2\Delta ^2(x)\cos ^2(2(\theta
-\alpha ))}},\qquad 
\]

\[
\Psi (x)=\frac{\sqrt{2}\cos (2(\theta -\alpha ))\Delta (x)}{\sqrt{\omega
^2+2\Delta ^2(x)\cos ^2(2(\theta -\alpha ))}}. 
\]

Here prime denotes the derivative with respect to the coordinate $x$. As is seen
from the solution (\ref{Eq.12}), the anomalous Green's function $\Phi _{+}$ at $x=0$ is
proportional to the sum of three terms with different angular symmetry. Two
of them are the isotropic part $\left\langle \Phi _{+}(0)\right\rangle $ and
the term with the $d$-wave symmetry $\Psi (0)$, while the third term is
proportional to the product $\Delta ^{\prime }(0)\left| \cos \theta \right|
\cos (2(\theta -\alpha ))$. The latter term is a source for nucleation of a nonzero 
$s$-wave component of  $\Phi _{+}$. 

\begin{figure}
\par
\begin{center}
\mbox{\epsfxsize=\hsize \epsffile{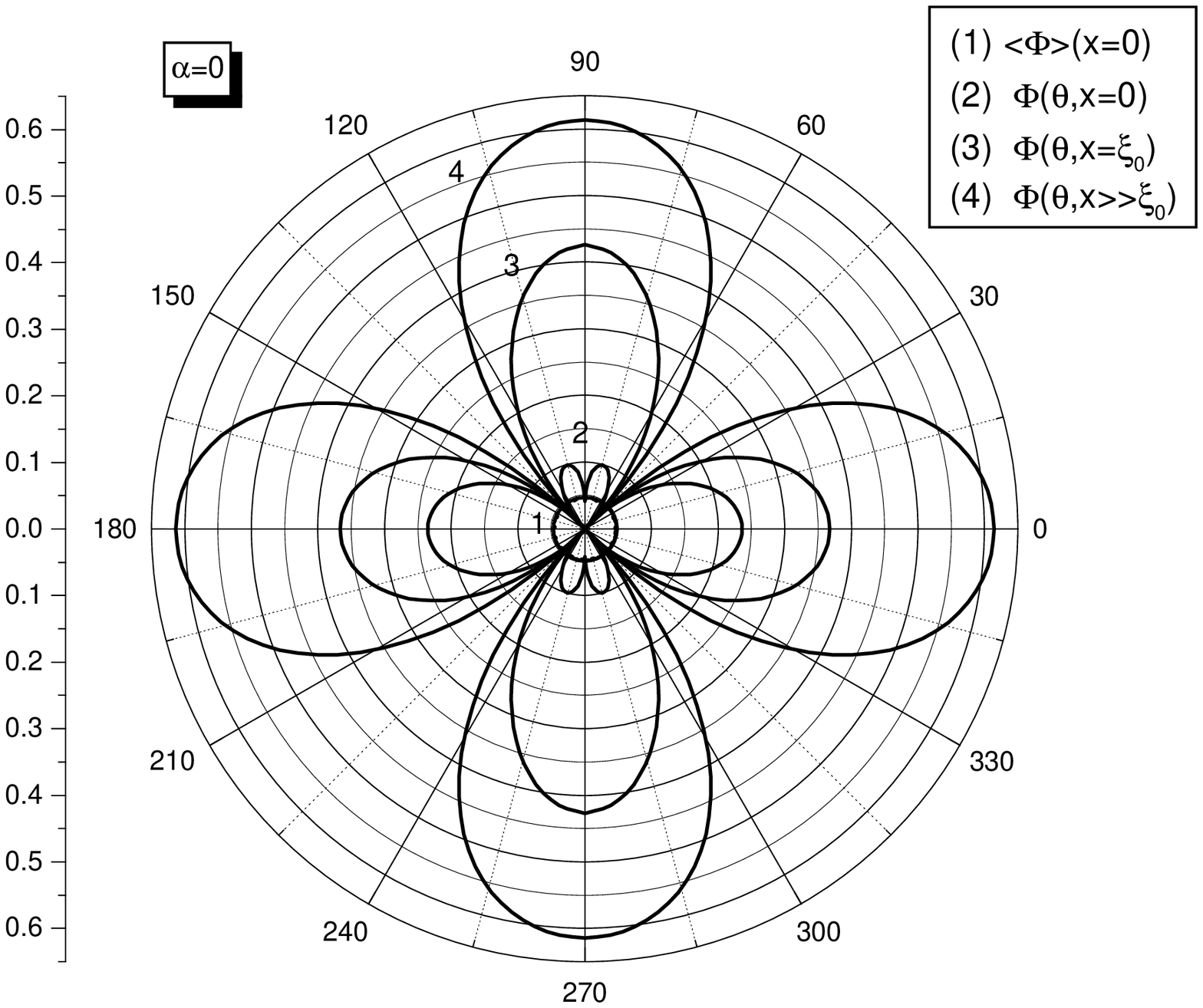}}
\mbox{\epsfxsize=\hsize \epsffile{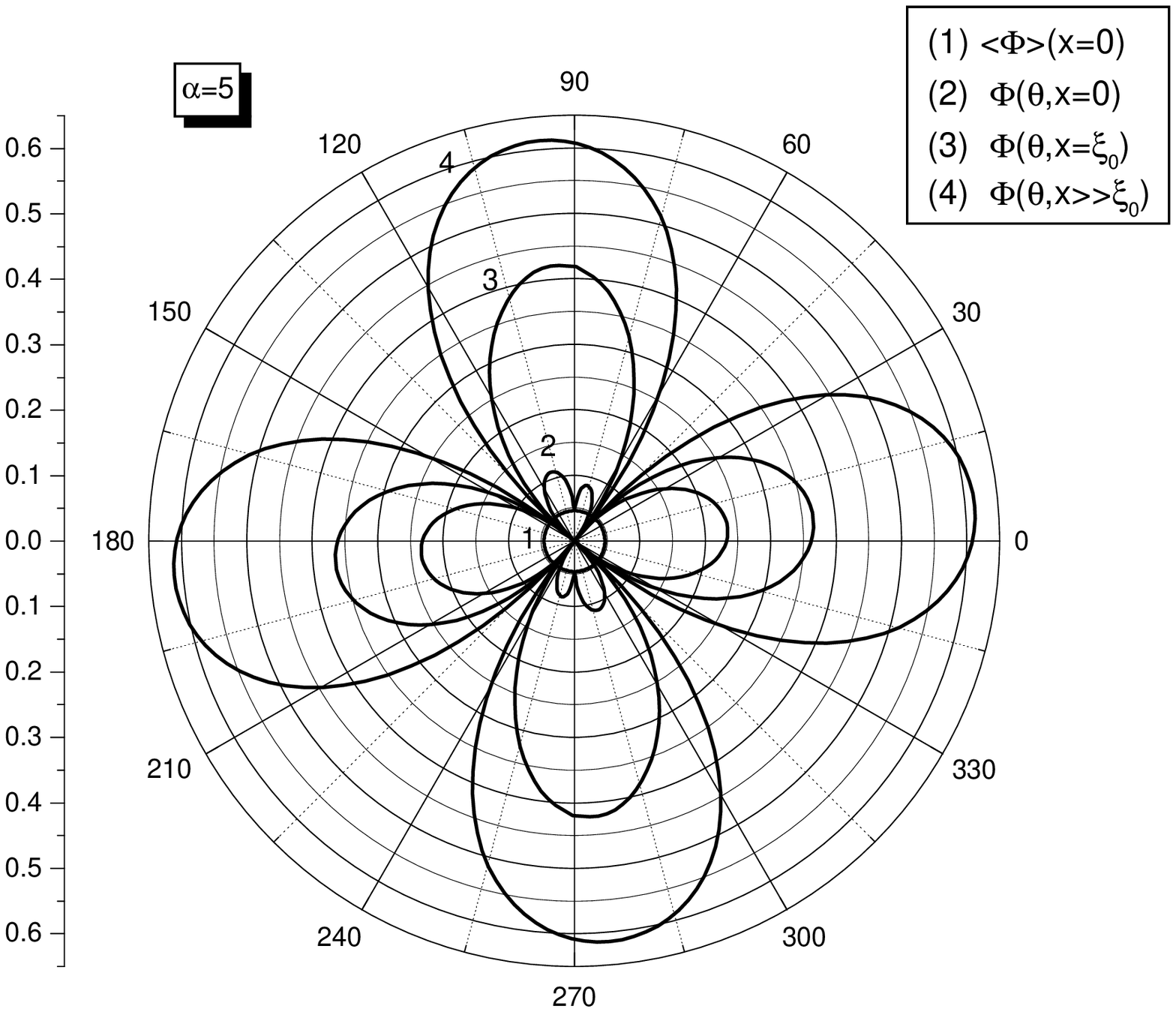}}
\end{center}
\caption{Angular dependencies of $\Phi _{+}(x)/\pi T_c$ at different distances from
the interface at $T=0.7T_c$. (a) Misorientation angle $\alpha =0$,
(b) $\alpha = 5^0$}
\end{figure}

Note that according to Eq.(\ref{Eq.12}) the solution $\Phi _{+}$ at $x=0$ would have
pure $d$-wave symmetry in the approximation of a spatially independent pair potential 
$\Delta$. The reason is that in this case the characteristic length $\kappa(\theta)$ 
of spatial variation of $\Phi _{+}$ cancels out from the solution Eq.(\ref{Eq.12})
for $\Phi _{+}(x)$, since $\kappa(\theta)$ is present both in Eq.(11) 
and in the boundary condition Eq.(13). At the same time, in the selfconsistent approach 
a nonzero angle-averaged value $\left\langle \Phi _{+}(x)\right\rangle $ appears at
the interface, since the above cancellation is incomplete in the presence of a pair 
potential gradient.

As suggested in \cite{Rainer1,Bah}, an $s$-wave component of the order 
parameter may nucleate at the surface of a $d$-wave superconductor if there
is a subdominant bulk pairing interaction in the $s$-wave channel. We 
have demonstrated above that the nonzero $s$-wave component 
$\left\langle \Phi _{+}(x)\right\rangle $ is localized near the rough 
interface even in the absence of a subdominant interaction in the bulk.

It is worth mentioning that since a subdominant $s$-wave pairing interaction 
is not included in the present model, the pair potential $\Delta $ still has 
pure $d$-wave angular structure everywhere in the $d$-wave region.
For the same reason there is no source for the phase shift between $s$- and 
$d$-wave components of $\Phi _{+}(x)$, thus the surface $d+is$ state which 
breaks time-reversal symmetry should not occur in the case considered.

In the general case of arbitrary $\kappa $\ values the problem was solved
numerically. The isotropic function $\left\langle \Phi _{+}(0)\right\rangle $\ 
and the spatially dependent pair potential $\Delta (x)$\ were calculated
by iterating the equations (11), (12) making use of the boundary condition 
(\ref{Eq.11}) and the selfconsistency equation (\ref{Eq.4}). The results of 
numerical calculations shown in Figs.1-3 confirm the considerations presented above.

Fig.1 shows the angular dependence of $\Phi _{+}(x)$ far from the boundary ($%
x\gg \xi _0$), as well as at $x=\xi _0$ and $x=0$, for two different orientations 
$\alpha $ of the $a$ axis with respect to the interface normal. In both cases far from the
interface the angular distribution is typical for a $d$-wave superconductor.
At $x=\xi _0$ the positive lobe (horizontal) is suppressed stronger than the negative
one (vertical), since the characteristic length $\kappa (\theta )$ in the direction
perpendicular to the interface is small compared to $\kappa (\theta )$ in the 
direction parallel to the interface. Hence at $x\approx
\xi _0$ negative lobes of $\Phi _{+}(x)$ practically reach the local value $%
\Psi (x)$, while positive ones still do not. This difference leads to the
negative sign of the $s$-component $\left\langle \Phi _{+}(x)\right\rangle $
(see Fig.2). 

\begin{figure}
\par
\begin{center}
\mbox{\epsfxsize=\hsize \epsffile{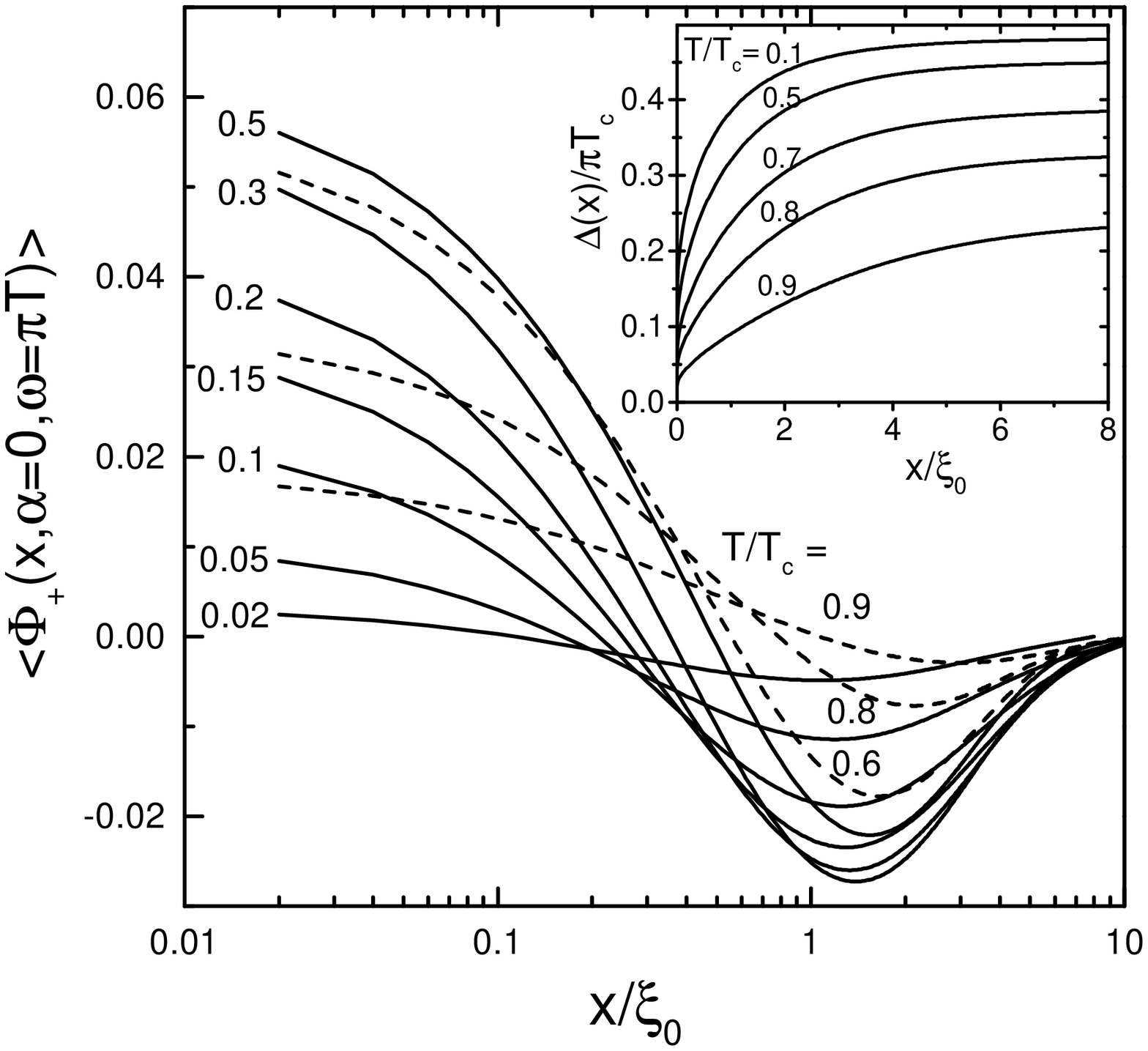}}
\end{center}
\caption{Spatial dependencies of the surface-induced $s$-wave component $%
\left\langle \Phi _{+}(x)\right\rangle /\pi T_c$ at various temperatures. Insert:
behavior of the pair potential $\Delta (x)/\pi T_c$ near the interface}
\end{figure}

In the vicinity of the interface $(x\leq 0.3\xi _0)$ the situation is 
just the opposite. In accordance with solution (\ref{Eq.12}), due to the
angular dependence of $\kappa (\theta )\propto $ $\left| \cos \theta \right| 
$ the negative lobes are suppressed stronger than the positive ones, the
function $\left\langle \Phi _{+}(x)\right\rangle $ changes sign to positive
and reaches its maximum at $x=0.$ 

Note that for quasiparticle trajectories parallel to the surface,
$\theta =\pi /2$, it follows from Eq.(\ref{Eq.12}) that $\Phi _{+}(0,\pi /2)=\Psi
(0,\pi /2)$, while $\lim_{\theta \rightarrow \pm \pi /2}\Phi _{+}(0,\theta
)=\Psi (0,\pi /2)+\eta .$ This discontinuity is the manifestation of the
simple fact that quasiparticles which propagate exactly parallel to the
interface have information about the disordered region only via the local
value of the pair potential, while for all other directions the direct interaction
between both regions takes place. However, this discontinuity at $\theta
=\pi /2$ does not contribute to the result of the angular averaging of $\Phi
_{+}$ in the boundary condition Eq.(\ref{Eq.11}).

\begin{figure}
\par
\begin{center}
\mbox{\epsfxsize=\hsize \epsffile{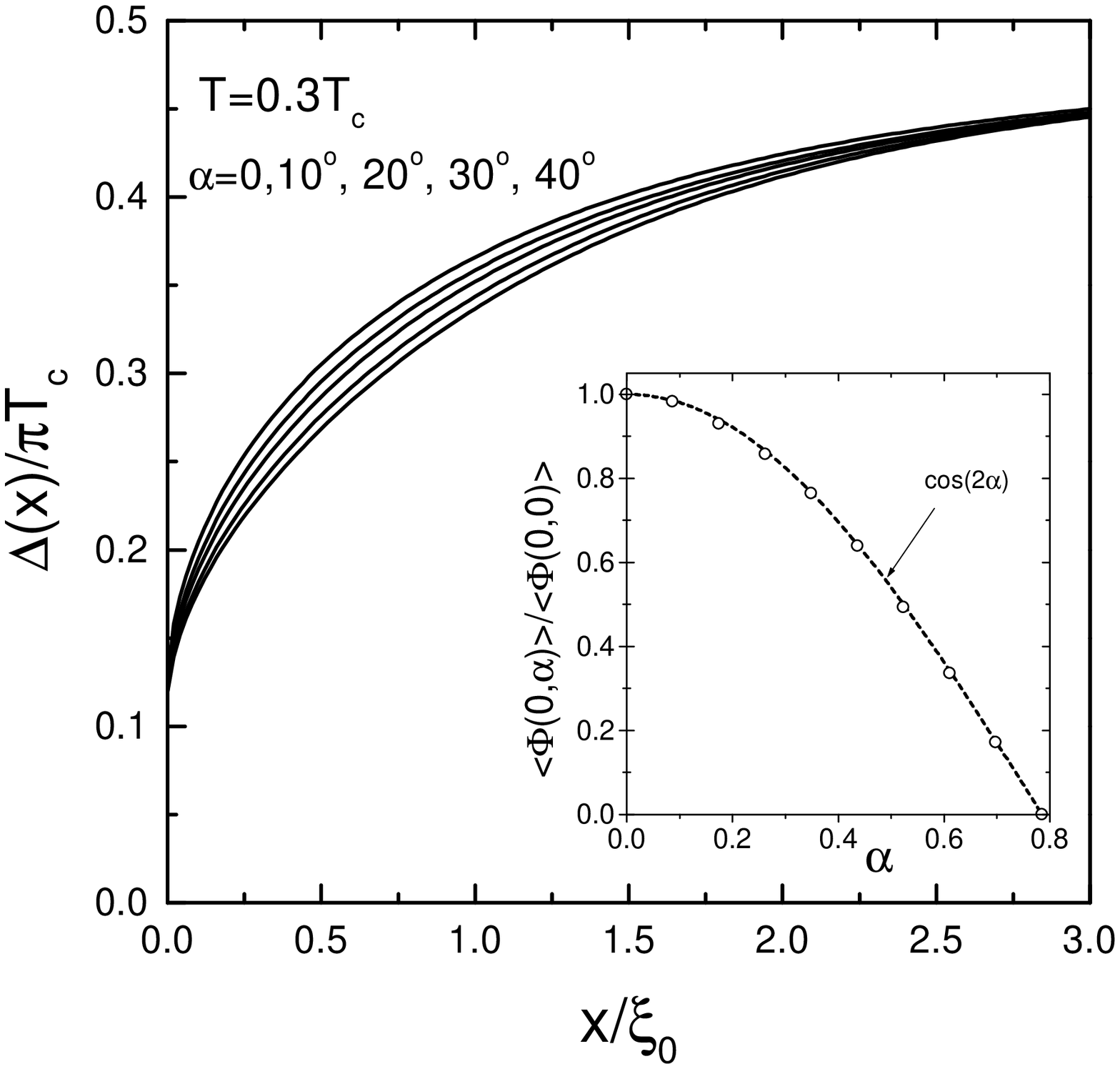}}
\end{center}
\caption{Behavior of the pair potential near the interface for different
misorientation angles $\alpha $. Insert: dependence of $\left\langle \Phi
_{+}(0)\right\rangle $ on $\alpha $.}
\end{figure}

Fig.3 shows the spatial variations of $\Delta (x)$ for different values of
the angle $\alpha $. As follows from Eq.(\ref{Eq.12}), the function $\Phi
_{+}(0,\theta )$ near the interface has a contribution proportional to $\left|
\cos \theta \right| \cos (2(\theta -\alpha ))$. This immediately leads to
the result that the amplitude of the $s$-component induced into the
disordered layer scales with misorientation angle $\alpha $ as $\left\langle
\Phi _{+}(0,\alpha =0)\right\rangle \cos 2\alpha .$ At $\alpha =\pi /4$ the
superconducting correlations are not induced into the disordered layer, i.e. 
$\left\langle \Phi _{+}(0)\right\rangle =0$. Further increase of $\alpha $
leads to a sigh change of the $s$-component. 

As is seen from Fig.3, these
qualitative considerations are in a good agreement with the results of exact
numerical calculations. In particular, for the $d_{xy}$ case ($\alpha =\pi /4$)
it follows that $\left\langle \Phi _{+}(0)\right\rangle =0$. At the same time, 
it is worth mentioning that the pair potential at the interface, 
$\Delta (0,\alpha =\pi /4)$, is nonzero, in contrast to the case of a 
specular reflecting boundary when $\Delta$ at $\alpha =\pi /4$ vanishes.
In the case considered of diffusive surface scattering there is no symmetry 
requirement for the vanishing of $\Phi _{+}(0,\alpha =\pi /4)$.

In the whole temperature range the amplitude of the $s$-wave
component $\left\langle \Phi _{+}\right\rangle $ induced into the disordered
layer is an order of magnitude smaller compared to the amplitude
of the order parameter in the bulk superconductor (see Fig.2). That means that $%
\left\langle g(0)\right\rangle $ is close to unity for all temperatures.
Thus, taking into account that $\left\langle
g(0)\right\rangle $ is independent of the Matsubara frequencies and that $\xi
\Phi _{+}^{\prime }(0)\approx \Phi _{+}(0)$, we obtain from the boundary condition 
Eq.(\ref{Eq.11}) that at low temperature $\left\langle \Phi
_{+}(0)\right\rangle $ $\propto \omega$ for $\omega \leq \Delta $. As soon
as $\omega $ exceeds the value of $\Delta $, the function $\left\langle \Phi
_{+}(0)\right\rangle $ behaves as $\left\langle \Phi _{+}(0)\right\rangle $ $%
\propto \omega ^{-2}$. The density of states 
$N(\varepsilon ) = N(0) Re \left\langle g(0,\varepsilon =i\omega )\right\rangle $, 
where $\left\langle g\right\rangle =\sqrt{1-\left\langle \Phi _{+}\right\rangle ^2}$
and $N(0)$ is the normal state density of states. Therefore it follows from
the property $\left\langle \Phi _{+}\right\rangle $ $\propto \omega $ at small 
$\omega $ that the density of states $N(\varepsilon=0)/N_0=1$, i.e.  
there is a gapless superconducting state in the disordered layer. 

\begin{figure}
\par
\begin{center}
\mbox{\epsfxsize=\hsize \epsffile{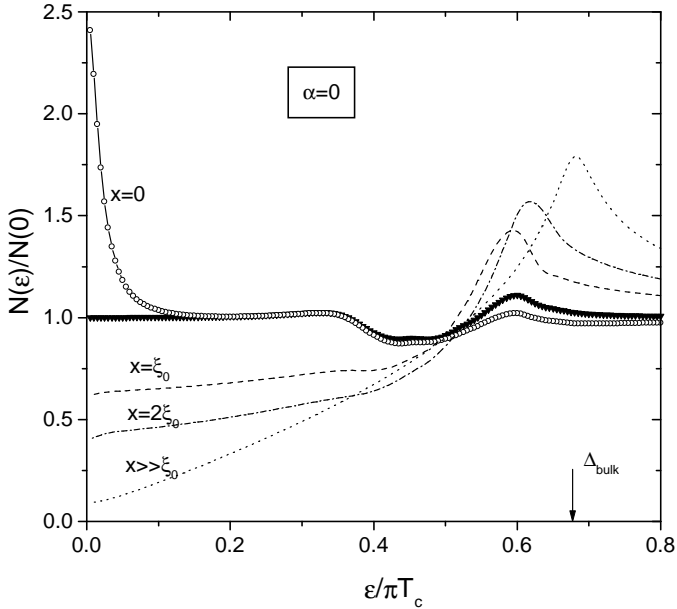}}
\end{center}
\vspace{0.5cm}
\caption{The densities of states at $T=0.1T_c$ in the disordered layer
(solid triangles) and in the $d$-wave region at $x=0$ (open circles), $x=\xi _0 
$ (dashed), $x=2\xi _0 $ (dash-dotted) and in the bulk (dotted)}
\end{figure}

To demonstrate this behavior explicitly we have calculated 
the density of states by numerical integration 
of Eqs.(11)-(13) on the real energy axis making the substitution 
$\omega =-i\varepsilon $ in these equations. The results of calculations of the
normalized low-temperature density of states $N(\varepsilon)/N(0)=1$
at $T=0.1T_c$ and $\alpha =0$ are presented in Fig.4.
To take into account an inelastic scattering we have introduced the complex energy 
$\widetilde{\varepsilon }=\varepsilon +i\gamma $ with $\gamma =0.05T_c$.

As is seen from Fig.4, the density of states in the disordered layer 
is gapless and has a rather
weak and broad peak at an energy below the maximum value of the bulk pair
potential $\Delta_{bulk}$. This peak is a signature of the Andreev bound
states at finite energies. Note that there is no midgap (zero-energy) peak in
the density of states in the disordered layer since there is no sign change of the 
order parameter in this region. At the same time the density of states at the 
surface of the $d$-wave region (at $x=0$) exhibits a sharp zero-energy peak.
In accordance with the known results for specular interfaces \cite
{Kashiwaya,Barash1,Rainer1} this peak is due to the midgap states in the
surface of the $d$-wave region .

An important difference between specular and rough interfaces is that in the former
case the midgap states occur only at nonzero values of the misorientation angle
$\alpha \neq 0$, whereas in the latter case these states occur even for $\alpha =0$. 
This result has a simple physical interpretation. Due to the presence of the
disordered surface layer incident and reflected quasiparticle trajectories are uncorrelated. 
Therefore for any $\alpha $ there is a finite probability for an incident quasiparticle
to experience a sign reversal of the pair potential upon reflection. Then the
averaging over incoming trajectories yields the midgap states and as a result
the nonvanishing zero-energy peak in the surface density of states appears.

\section{Conclusions}

In conclusion, the proximity effect between a $d$-wave superconductor and a 
disordered surface layer is studied theoretically in the regime of strong disorder.
The boundary conditions for 
the Eilenberger equations are derived at the interface between the clean and the
disordered regions. It is shown that superconducting correlations in the disordered 
layer do not vanish in the limit of small electronic mean free path and the 
isotropic superconducting state is induced in such a layer. The crossover from 
this state to the $d$-wave pairing state in the bulk is studied by solving 
the Eilenberger equations. 

The quasiparticle density of states in the disordered layer is gapless
and shows a broad peak at finite energies, while the density of states at the 
surface of the $d$-wave region exhibits a zero-bias peak due to sign reversal 
of the order parameter. This peak is fully smeared out in the disordered 
layer. 

The above phenomena have important consequences for the description of rough HTS 
interfaces and Josephson junctions, in particular for SNS junctions with a normal 
metal interlayer. These effects will be discussed elsewhere.

{\bf Acknowledgments}. We would like to thank J.Aarts, G.J.Gerritsma,
Yu.Nazarov and H.Rogalla for helpful discussions. This work is supported 
in part by INTAS Grant 93-790ext and 
by the Program for Russian-Dutch Research Cooperation (NWO).

\end{document}